\documentclass{article}  
\usepackage{arxiv}
\usepackage{graphicx}
\usepackage{amsmath}
\usepackage{subfloat}
\usepackage{subfig}
\usepackage[utf8]{inputenc} 
\usepackage[T1]{fontenc}    
\usepackage{url}            
\usepackage{booktabs}       
\usepackage{amsfonts}       
\usepackage{nicefrac}       
\usepackage{microtype}      
\usepackage{lipsum}
\usepackage{array}
\title{A Deep Learning Scheme for Efficient Multimedia IoT Data Compression}

\author{
  Hassan N. Noura  \\
Univ. Bourgogne Franche-Comté (UBFC),\\ FEMTO-ST Institute,\\ France\\

   \And
   Ola Salman\\
   American University of Beirut,\\
   Electrical and Computer Engineering Department,\\ Beirut 1107 2020, Lebanon\\
   
    \And
Raphaël Couturier\\
Univ. Bourgogne Franche-Comté (UBFC),\\ FEMTO-ST Institute,\\ France\\

}

\begin{document}
\maketitle
\begin{abstract}
Given the voluminous nature of the multimedia sensed data, the Multimedia Internet of Things (MIoT) devices and networks will present several limitations in terms of power and communication overhead. One traditional solution to cope with the large-size data challenge is to use lossy compression. However, current lossy compression schemes require low compression rate to guarantee acceptable perceived image quality, which results in a low reduction of the communicated data size and consequently a low reduction in the energy and bandwidth consumption. Thus, an efficient compression solution is required for striking a good balance between data size (and consequently communication  overhead) and visual degradation. In this paper, a Deep-Learning (DL) super-resolution model is applied to recuperate high quality images (at the application server side) given as input degraded images with a high compression ratio (at the sender side). The experimental analysis shows the effectiveness of the proposed solution in enhancing the visual quality of the compressed and down-scaled images. Consequently, the proposed solution reduces the overall communication overhead and power consumption of limited MIoT devices.
\end{abstract}


 

\section{Introduction}
Recently, the advancement in information technology has enabled new emerging networks such as the Internet of Things (IoT)~\cite{goelreview}. IoT is meant to connect different types of physical objects and devices~\cite{wang2017multimedia}. Once connected, these devices can transmit data enabling innovative applications. Given the massive scale and heterogeneous nature of IoT, several performance threats will emerge~\cite{yaacoub2020securing}, many of which remains widely unexplored in existing works. Indeed, owing to its scale and heterogeneity, IoT is more constrained in terms of communication and power resources compared to conventional communication networks~\cite{tanwar2019multimedia}. \\

Nowadays, IoT is witnessing an extensive development with the emergence of new wireless technologies and new types of devices. In this context, the availability of cheaper hardware, as CMOS cameras and microphones, has enabled the development of large-scale Multimedia IoT (MIoT) networks~\cite{wang2017multimedia,nauman2020multimedia,al2020survey}. Therefore, a set of IoT applications have been developed, requiring the transmission of a large amount of multimedia data such as images, audios, or videos.

\subsection{Problem Formulation}
Multimedia data differs from standard scalar data in that it is larger in scale and could be delivered with real-time constraints. These features impose a range of conditions on the operation of MIoT networks, including:
\begin{itemize}
    \item A wide transmission bandwidth is required to facilitate multimedia streams, which is a hard constraint in practice.
    \item The supply of additional or powerful Central Processing Unit (CPU) capabilities to MIoT devices to process multimedia data.
    \item Given the constraints of the MIoT devices in terms of computation and communication resources, new multimedia data transfer algorithms are required.  
\end{itemize}

Thus, there is a growing need to increase the efficiency and reliability of the MIoT transmission. Data compression is a main function employed for communicated data size reduction. The compression efficiency can be improved by reducing more and more the size of the compressed multimedia data, which results in reducing the communication, computation, and time overhead. This can be achieved by a hard lossy compression with a high compression ratio. However, this comes at the cost of the content quality.  For example, in the JPEG image compression, the image quality is inversely proportional to the compression ratio. To preserve the visual quality of multimedia contents, a low compression ratio is required.  Thus, there is a need of new data processing techniques that optimize the trade-off between the communicated data size (or compression ratio) and the content quality.


\subsection{Motivation}
MIoT devices might have limited computation and communication resources, in addition to power constraints as they are mostly battery-driven. Therefore, the main target of this work is to preserve the MIoT devices' resources and lifetime while transmitting voluminous data, given that data transmission is one of the most energy-consuming functions~\cite{jan2017balanced}. In this paper, an efficient multimedia data-reduction solution is presented to handle the communication and power constraints of limited MIoT devices.  \\

Hence, reducing the communicated data size between MIoT devices and application server(s) should be one of the ultimate goals of any technique aiming at reducing the MIoT bandwidth and latency overhead. However, existing multimedia compression algorithms still present deficiency in meeting the balance between compression ratio and image quality. Recently, Deep Learning (DL) was employed in the computer vision domain to perform hard tasks that were undoable with traditional machine learning algorithms~\cite{li2018learning,tang2017enabling}. In this context, DL has been applied for image denoising and super-resolution~\cite{tian2018deep,tian2020deep,bai2020survey,wang2020deep}, which can help in alleviating the effect of the hard visual degradation. Thus, our aim is to employ existing compression techniques like JPEG and BPG with a high compression rate while restoring high image quality at the application server by applying a super-resolution \& denoising DL approach. 


\subsection{Contributions}
To the best of our knowledge, this is the first work designing a DL-based solution to handle the MIoT intrinsic limitations. This approach integrates the data reduction at the MIoT devices (by applying data compression with high compression rate) with the visual content enhancement and super-resolution at the application server(s) (by employing a DL model). It should be noted here that any denoising \& super-resolution DL model, that can achieve good results, can be applied~\cite{wang2020deep}.\\


In this paper, a novel approach is proposed to reduce the amount of communicated multimedia data between MIoT devices and application server(s). The quality of the received multimedia contents can be enhanced by using an efficient denoising super-resolution DL model such as Residual Dense Network (RDN)~\cite{zhang2018residual}. This paper presents two variants of the data reduction solution. While both variants employ data compression, the second variant applies the down-scaling operation to reduce the image size before compression, and consequently, more data reduction can be achieved compared to the first variant at the cost of higher visual degradation. \\

Mainly, the proposed approach is designed to satisfy the following objectives: 
\begin{enumerate}
    \item Avoiding additional computation complexity at MIoT devices.
    \item Reducing the amount of communicated data to reach lower communication overhead and to reduce the latency in addition to the transmission energy consumption. 
    \item Preserving (or enhancing) the perceived visual quality at the application server(s).
    \item Designing a flexible denoising and super-resolution solution to be applied with different image qualities (or compression ratios) and down-scaling sizes, depending on the MIoT devices and applications requirements. 
\end{enumerate}

 
\begin{figure*}[!ht]
\centering
\includegraphics[scale=0.5]{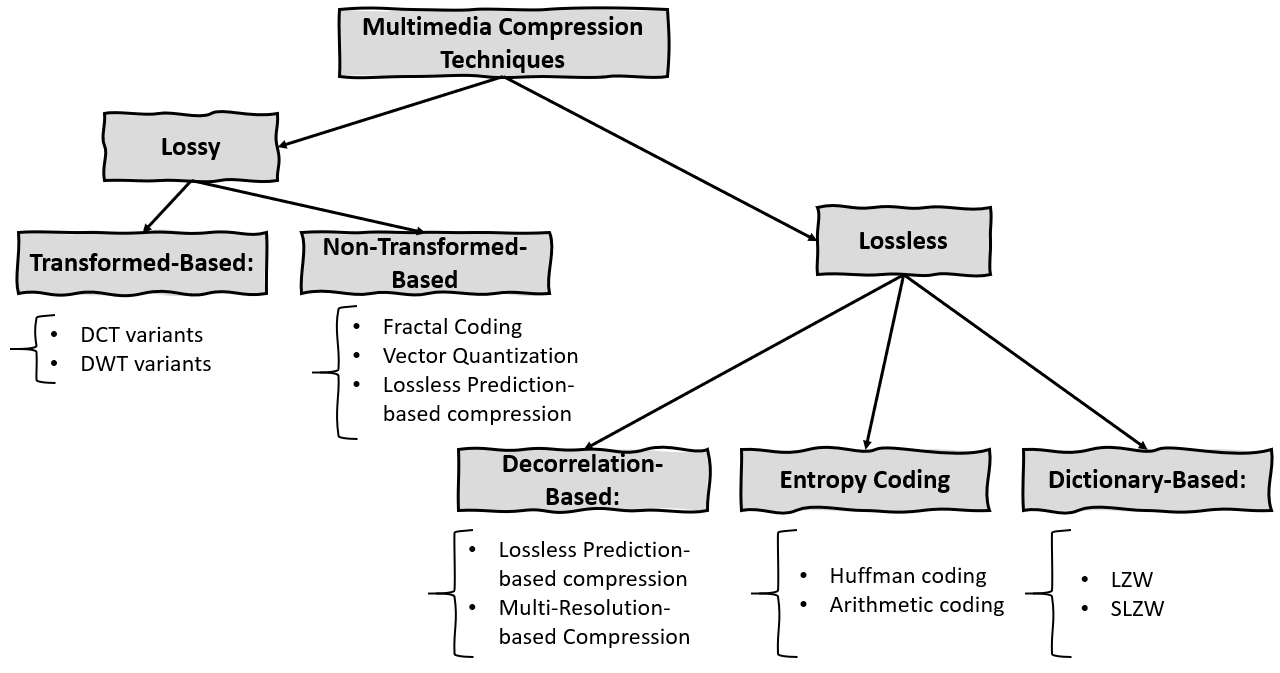}
\caption{Taxonomy of multimedia compression}%
\label{fig:compressionRatioSize2}%
\end{figure*}

\subsection{Organization}
The rest of this paper is organized as follows: Section~\ref{sec:background} reviews the related work, especially the recent DL denoising and super-resolution approaches. Then, the proposed system model is presented and described in Section~\ref{sec:proposedSolution}. Then, Section~\ref{sec:performance} details the experimental setup and the obtained results. Section~\ref{sec:discussion} discusses the obtained results and presents future research directions. Finally, Section~\ref{sec:conclusion} concludes the paper.

\section{Background \& Related Work}
\label{sec:background}
In this section, we start by briefly describing the well-known image compression algorithms, which are JPEG and BPG. After this, a set of recent DL-based image denoising and super-resolution approaches are discussed. 

\subsection{ Image Compression}
The image compression techniques can be divided into two classes: lossless and lossy (as illustrated in \figurename~\ref{fig:compressionRatioSize2}). The lossless compression techniques preserve better the visual content compared to the lossy ones since the decompressed image with lossless compression is the same as the original one. However, the lossless compression results in larger compressed data size compared to the lossy one. In general, the lossy compression is used to reduce the bandwidth overhead in existing networks. In fact, the compression ratio is inversely proportional to the image quality. Using a low compression ratio results into a high perceived image quality and vice versa.\\

Existing lossy compression techniques convert data from the spatial to the frequency domain by applying a spatial-frequency transformation such as the Discrete Cosine Transform (DCT2) used in JPEG~\cite{wallace1992jpeg} and BPG or Discrete Wavelet Transform (DWT2) in JPEG 2000~\cite{skodras2001jpeg}. On the other hand, DCT2 is also used in the well-used lossy video compression standards such as MPEG 1/2, AVC/H.264, MPEG-4, and HEVC. After the spatial-frequency transformation, the frequency values are quantized (according to the frequency level) and ordered, then they get through the Huffman or arithmetic encoding process. Given that the human eye is good in remarking changes in low frequency regions while disregarding the changes in the busy patterns regions, the lossy compression techniques represents the less important data with the high frequency components, while the most important information are represented by the lowest frequencies. Consequently, the lossy compression targets to eliminate more data from the high frequency components than from the low frequency ones. Therefore, the frequency coefficients have different importance levels and consequently they are quantized according to their importance and the desired compression ratio. This can lead to different visual effects. A low compression ratio means that only high and middle-frequency coefficients are quantized hardly compared to low-frequency coefficients.  \\

In the following, JPEG and BPG image compression algorithms will be described.

\subsubsection{JPEG Compression Standard}
\label{sec:jpeg}
JPEG~\cite{wallace1992jpeg} is a lossy image compression standard that was developed by the Joint Photographic Experts Group (JPEG) committee. It is based on the concept of frequency transform coding, where DCT is used as frequency transformation on each image block ($8\times 8$). 

The JPEG compression process can be divided into five main steps, which are:
\begin{enumerate}
    \item \textbf{Pre-processing}: 
    In this step, a color transformation is applied to convert the pixels of the $R$ (Red), $G$ (Green), and $B$ (Blue) components into $Y$ (Luminance), $C_b$ (Blue-difference Chroma) and $C_r$ (Red-difference Chroma) components. Then, the chroma sub-sampling process is applied.
    \item  \textbf{Discrete Cosine Transform (DCT)}: 
    After preprocessing, the output matrices are divided into blocks (sub-matrix) of equal size (e.g 8$\times$ 8). Then, a frequency transformation is applied to each image block. This transformation decomposes each block into high, middle, and low-frequency sub-bands (DCT coefficients), resulting into 64 coefficients for each block.  Each obtained coefficient carries distinct information of the transformed signal. The first DCT coefficient is called the DC coefficient, which carries the average intensity of the transformed block (64 elements). The rest DCT coefficients are called AC coefficients. Therefore, DC consists of the most important lowest frequency. DC and first AC coefficients represent the low frequencies, while other AC coefficients represent middle and high frequencies.
    \item  \textbf{Quantization}: 
    In this step, the DCT coefficients are quantized using a quantization matrix. The quantization matrix depends on the compression ratio (image quality). In this step,  many of the AC coefficients (middle and high frequencies) will be ignored. The quantization and rounding step determines the number of DCT coefficients that will be rounded to zeros.
    \item \textbf{Zigzag and run-length ordering} 
    \item \textbf{Encoding}: Finally, each compressed block is encoded using the Huffman entropy encoding.
\end{enumerate}

\subsubsection{BPG Compression Standard}
\label{sec:BPG}
The Better Portable Graphics (BGP) is an image format, presenting several advantages compared to JPEG~\cite{BPGIm8857370:online,albalawi2015hardware}. It achieves a lower visual degradation compared to JPEG, with the same compression ratio. In addition, BPG can be applied for lossless compression, provides the animation, and supports multiple color spaces (gray-scale, YCbCr, RGB, YCqCo) as well as chroma formats.\\

The BPG video encoder is based on HEVC~\cite{sze2014high}, which replaces H.264, a well-known video compression standard~\cite{wiegand2003overview}. HEVC presents good compression efficiency and it is considered as a major advance in the video compression domain. BPG as JPEG uses an 8$\times$ 8 block as the basic coding unit, and DCT as the frequency transformation mechanism. Besides, BPG can use the Discrete Sine Transform (DST) instead of DCT, reaching better performance compared to JPEG at the cost of higher computational overhead. Besides, many multimedia compression standards use the DCT transform on two dimensions (DCT-2D) with a small coding unit to reduce the overall time complexity. 

\subsection{DL-based Image Restoration Approaches}
The image restoration challenge can be considered as a very challenging task since the image degradation process is mostly irreversible. Recently, a set of DL-based models were presented to tackle this challenge and they are listed and described briefly in the following. \\

The current DL-based models have demonstrated a high-efficiency in extracting patterns from high dimensional data (images). These models can achieve promising results in several image restoration tasks such as image super-resolution, and image denoising. In this context, residual networks, a type of DL architectures, have been widely adopted for image super-resolution. A CNN-based super-resolution architecture has been proposed in~\cite{dong2014learning,dong2016accelerating}. This architecture is based on mapping the low-resolution parts of an image to their corresponding interpolated high-resolution parts. This architecture has been later enhanced by adding more layers and sharing weights. Introducing residual layers and adding substantial layers, Kim et al. present VDSR~\cite{kim2016accurate} and then add recursive connections with a deeper network with weights sharing in DRCN~\cite{kim2016deeply}. Tai et al. propose a similar architecture based on recursive blocks in DRRN~\cite{tai2017image} and then add memory to these blocks in Memnet~\cite{tai2017memnet}. A common characteristic of the residual-based architectures is adding skip connections between layers. This was applied in SRResNet~\cite{ledig2017photo} and EDSR~\cite{ledig2017photo} for designing efficient super-resolution networks. However, the main limitation of such networks is the reuse of features in just one layer per block. DenseNet was proposed permitting subsequent reuse of the features. Concatenating the layers instead of the summation in ResNet, DensNET achieves better performance. In this context, Tong et al. propose SRDenseNet~\cite{tong2017image} for image super-resolution by removing the pooling layers from DensNet.  Similarly, Zhang et al. use the dense residual blocks in RDN~\cite{zhang2018residual}. Haris et al. propose DBPN~\cite{haris2018deep}, which is constructed by iterative up and down sampling blocks restricting the block information flow during propagation. To guarantee a better flow of information between blocks, DBDN was proposed in~\cite{wang2018deep} enabling the reuse of the local and high-level information by adding intra-block and inter-block dense connections. More recently, RGSR was proposed as a two-step image super-resolution scheme having as input compressed images~\cite{li2021rgsr}. \\


In this paper, the Residual Dense Network (RDN)~\cite{zhang2018residual} model is used as a denoising and super-resolution model to be applied for recovering compressed images with low image quality (high compression ratio). To the best of our knowledge, this is the first work applying DL for MIoT compressed images recovery. The proposed solution aims to eliminate the hard visual degradation when MIoT devices use a high compression ratio with or without down-scaling to compress the transmitted images. The communicated data size depends on the selected image quality (compression ratio) in addition to the down-scaling factor (case of the second variant) that can be configured according to the devices and application requirements. JPEG and BPG codecs are considered as image compression methods, given that they can be implemented in the constrained MIoT devices (e.g. Raspberry PI, and Arduino).

\begin{figure*}[!ht]
\centering
\includegraphics[scale=0.42]{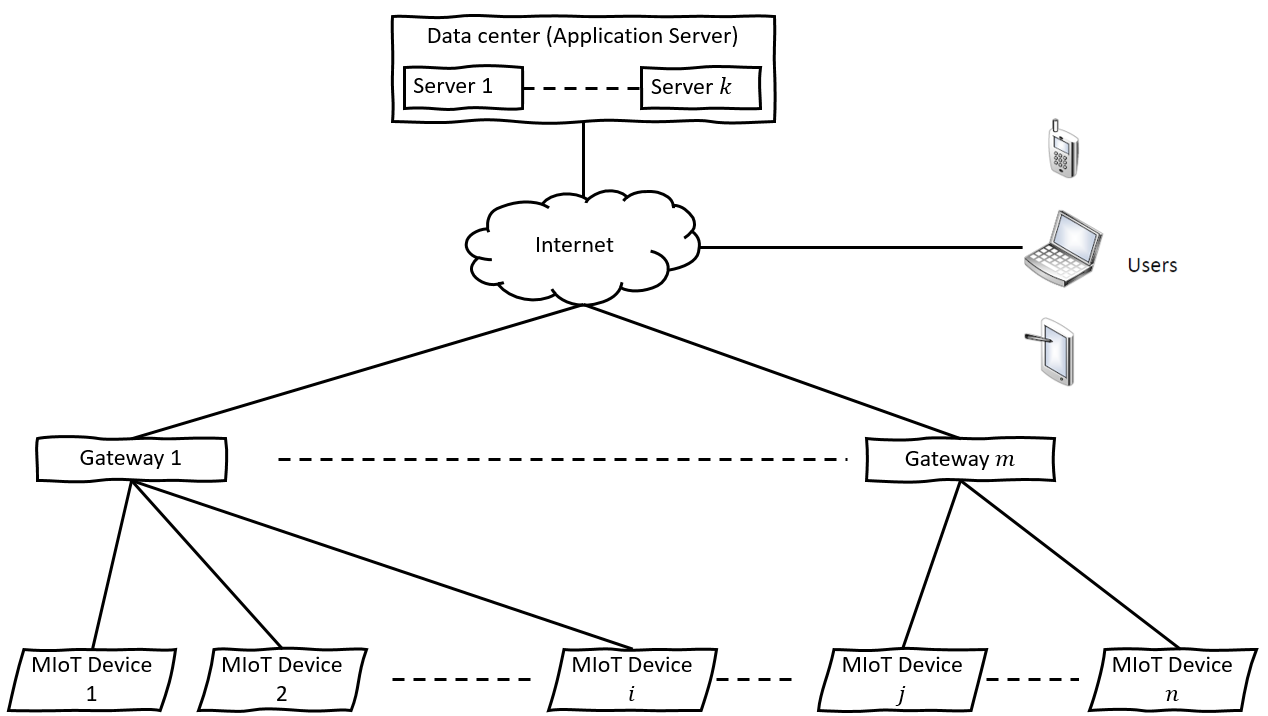}
\caption{An example of an MIoT network model: a set of $n$ MIoT devices, $m$ gateways, and the application server (or data center)}
\label{fig:MIoTmodel}
\end{figure*}

\begin{figure*}[!ht]
\centering
\subfloat[][First variant]{\includegraphics[scale=0.5]{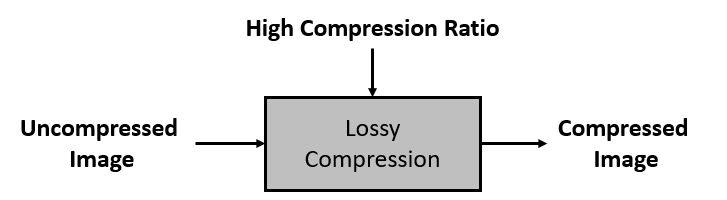}}\hfill
\subfloat[][Second variant]{\includegraphics[scale=0.5]{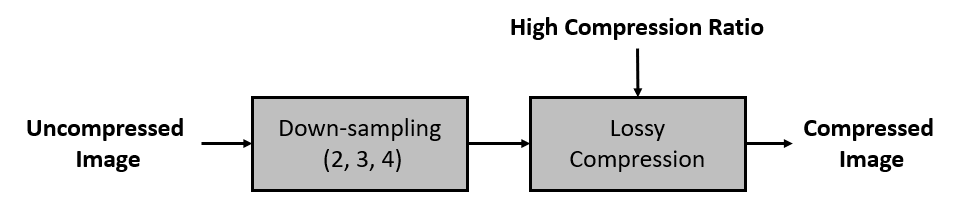}}\hfill
\caption{The proposed first (a) and second (b) variants}
\label{fig:var}%
\end{figure*}

\begin{figure*}[!ht]
\centering
\includegraphics[scale=0.5]{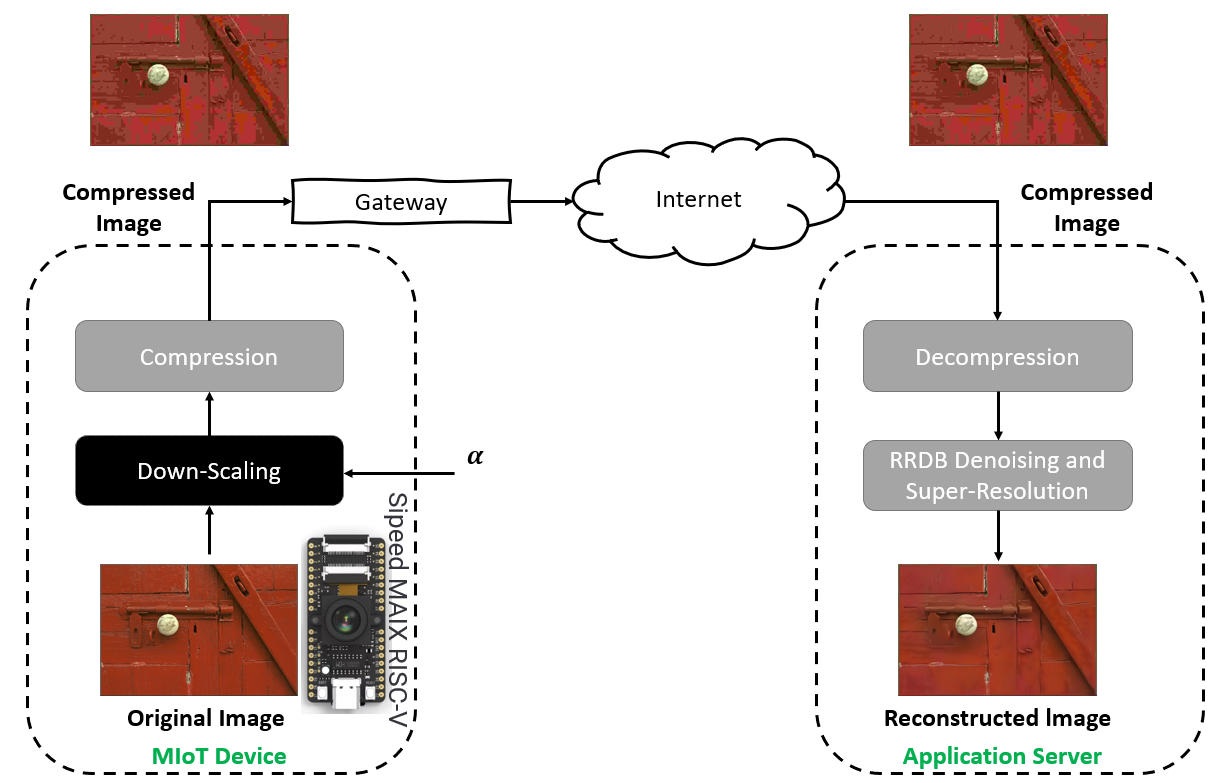}
\caption{Functional diagram of the proposed solution.}
\label{fig:modelsceme}
\end{figure*}

\section{System Model}
\label{sec:proposedSolution}
As shown in \figurename~\ref{fig:MIoTmodel}, MIoT consists of a large number of edge MIoT devices that can capture, process, and deliver multimedia data (e.g. images) to the data center or application server(s) through multi-hop or star wireless communications. Each MIoT device, after image acquisition, compresses the image and sends it to the application server. Then, the application server or data center decompresses the received image to recover the visual content. Given the large scale of the MIoT network and the large amount of communicated multimedia data, there is a real burden on the network bandwidth and MIoT devices' resources. Therefore, to reduce the required resources overhead, the compression algorithm realized on the MIoT devices should reduce the size of the transmitted image to the minimum possible to optimize the energy and bandwidth consumption.\\ 

Thus, our proposed solution aims to allow MIoT devices compressing the collected images with a high compression ratio to minimize the transmitted data size. Then, the visual contents of the received decompressed images can be enhanced at the application server(s) by employing a DL denoising \& super-resolution model such as the RDN model. Two variants of the proposed solution are considered (see \figurename~\ref{fig:var}): the first consists of having only compressed images sent by the MIoT devices, while the second variant considers sub-sampled and compressed images sent by MIoT devices.  In this context, our proposed system model, illustrated in \figurename~\ref{fig:modelsceme}, consists mainly of two steps:
\begin{enumerate}
    \item Using a lossy image compression with a high compression ratio, the MIoT devices send the compressed images to the application server. 
    \item Upon receiving the compressed images from MIoT devices, the application server (or data center) recovers their visual contents, which might present visual degradation, depending on the employed compression ratio. Then, the RDN model is applied to further enhance the quality of the decompressed images by reducing the compression noise effect. Let us indicate that, for the first variant of the proposed solution, the RDN  model plays the role of a denoiser, while, for the second variant, the RDN model ensures also the super-resolution property in addition to the denoising one since the images are down-scaled before being compressed. 
\end{enumerate}

In fact, the MIoT devices have different constraints, and using the same image quality for all applications is not practical. The image quality (and down-scaling factor for the second variant) depends on the target application requirements. Thus, the denoising and super-resolution model should be able to function with these different configurations (ensuring flexibility) towards effectively reconstructing the decompressed images. To this aim, the considered RDN model is trained (scaling or without scaling) with different image quality values to respond better to real-world scenarios. The proposed model aims to reduce the effect of lossy compression and down-scaling operations in contrast to existing applications of the RDN model, which aim to reduce the effect of other types of noise (e.g. Gaussian white noise). There is a range of targeted applications that can profit from the suggested approach, from military monitoring applications to automated assistance for older individuals, including advanced healthcare systems, home automation applications, etc. \\


\begin{figure*}[!ht]
\centering
\includegraphics[scale=0.5]{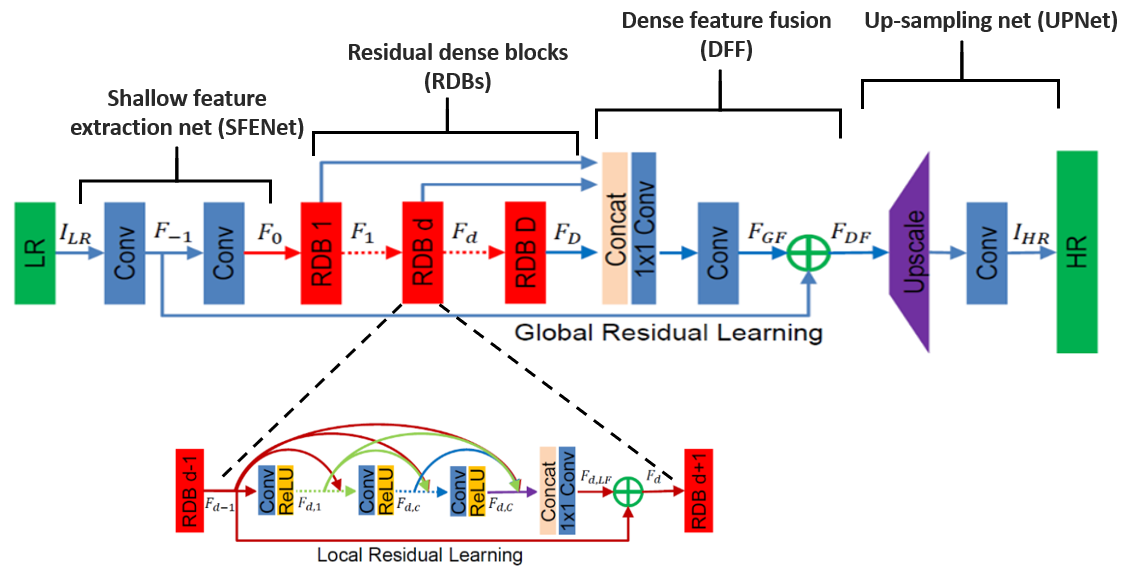}
\caption{Residual Dense Network (RDN) architecture~\cite{zhang2018residual}.}%
\label{fig:RDN}%
\end{figure*}

\subsection{RDN Architecture}
RDN consists mainly of four main components as illustrated in \figurename~\ref{fig:RDN}:
 \begin{enumerate}
     \item \textbf{ The Shallow Feature Extraction Net (SFENet)}: which consists of two convolutional layers, introduced to extract shallow features.
     \item \textbf{The Residual Dense Blocks (RDB)}: which take as input the extracted shallow features. 
     \item \textbf{The Dense Feature Fusion (DFF) layer}:  that fuses features from all the preceding layers, which are:
     \begin{itemize}
         \item The Global Residual Learning (GRL) layer: that processes the shallow feature-maps that represent the output of the first convolution layer of SFENet.
         \item The Global Feature Fusion (GFF) layer: that fuses features from all the RDB.
     \end{itemize}
     \item \textbf{The Up-sampling Net (UPNet)}
 \end{enumerate}


As illustrated in \figurename~\ref{fig:RDN}, the RDN architecture consists of $D$ identical Residual Dense Blocks (RDB). Each residual block contains $C=8$ convolution layers (with 3$\times$3 filters) and 1 convolution layer (with 1$\times$1 filters). In each residual dense block, there are $Ns$ shortcut connections, which represent an identity mapping that can help to solve the visual degradation problem appearing in stacked non-linear layers. Moreover, all layers use the Rectified Linear Unit (ReLU) as activation function to fit the residual dense mappings. The final convoluional layer has 3 output channels, given that the output high-resolution images can be colored, but also it can process gray images. For more information about the RDN architecture, readers can refer to~\cite{zhang2018residual}.\\

The benefit of the RDN model is that the hierarchy of all convolutional layers can be completely used. RDN consists of a $D$ Residual Dense Block (RDB), which can extract several local features (characteristics) through densely connected convolution layers. RDB creates direct links from previous RDB states to all current RDB layers, resulting into a Contiguous Memory (CM) mechanism. RDB's local feature fusion is used to acquire more effective features from previous and current local features in addition to stabilizing the learning of a wider network. The global feature fusion is employed to learn global hierarchical features jointly and adaptively in a comprehensive way after acquiring dense local features. Compared to DenseNet, RDN eliminates batch normalization and pooling layers. Pooling layers are removed from RDB since it could discard some pixel-level information, which is not preferable for image denoising and super-resolution. \\

The proposed solution applies the RDN model at the application server(s) to improve the image quality of the recovered decompressed images. Moreover, this model can also ensure the super-resolution property, which is beneficial for the second variant (that uses down-sampling). The purpose of RDN in the proposed approach is to learn mapping functions between the original uncompressed image $I$ and the compressed image $J$. For achieving this purpose, a set of denoising and super-resolution models were tested, and we select RDN since it can ensure acceptable performance. Any new image restoration model that can ensure better results can be used instead of RDN. The model's input image is $I=J+N$, where $J$ represents the compressed image and $N$ represents the reflecting block artifacts image introduced by lossy compression. The aim is to learn a residual dense mapping function between the original and the decompressed one.        

\begin{figure*}[!ht]
\centering
\subfloat[][JPEG]{\includegraphics[scale=0.55]{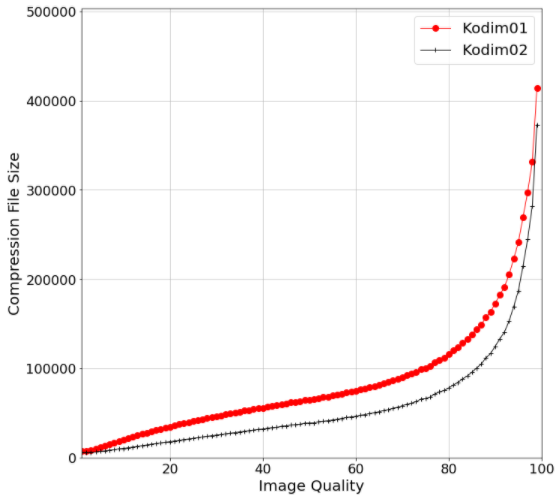}
}
\subfloat[][BPG]{\includegraphics[scale=0.6]{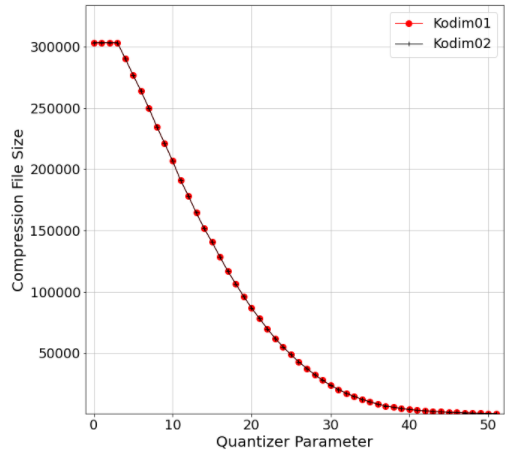}
}
\caption{Variation of the compression size versus compression image quality for two Kodak images with JPEG compressor.}%
\label{fig:SSIMversusQualityCmpression}%
\end{figure*}

\begin{figure*}[!ht]
\centering
\subfloat[][SSIM without down-scaling operation]{\includegraphics[scale=0.34]{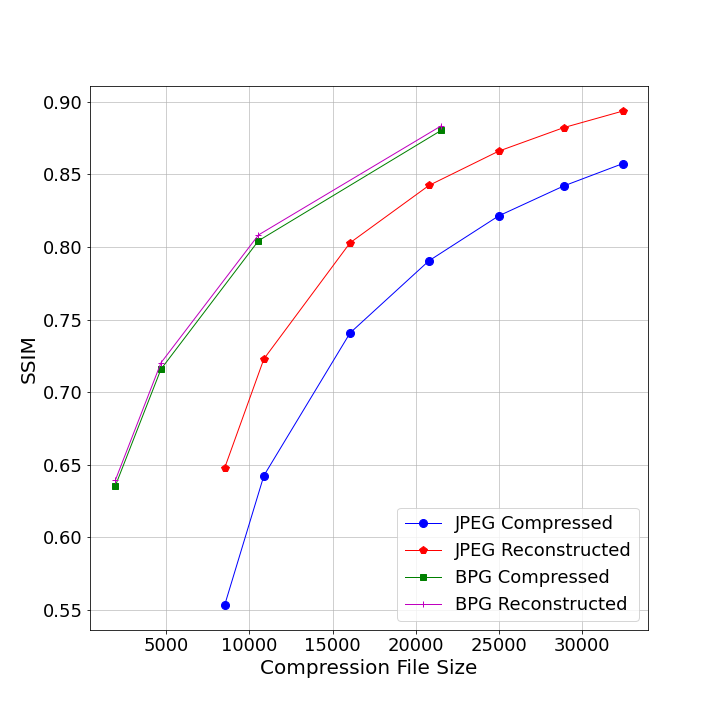}}
\subfloat[][SSIM with down-scaling operation]{\includegraphics[scale=0.34]{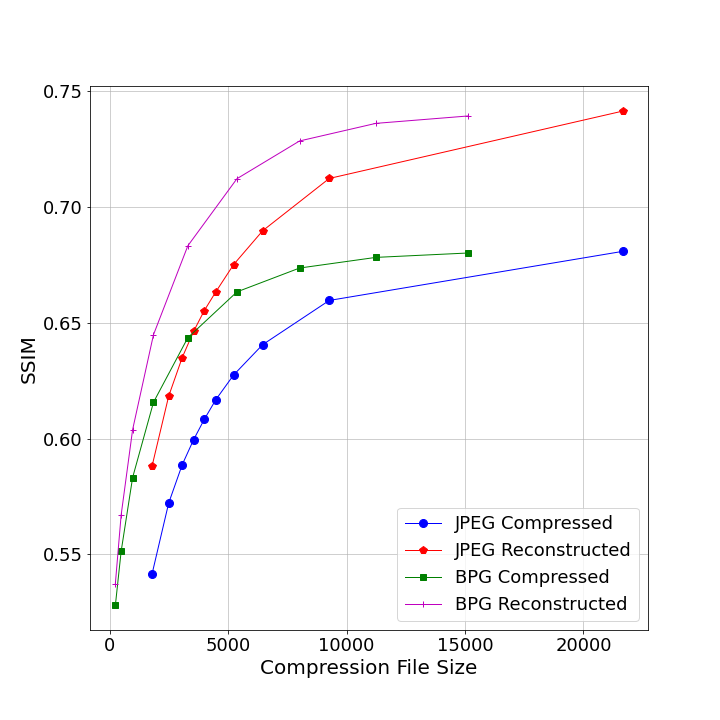}}\\
\subfloat[][PSNR without down-scaling operation]{\includegraphics[scale=0.34]{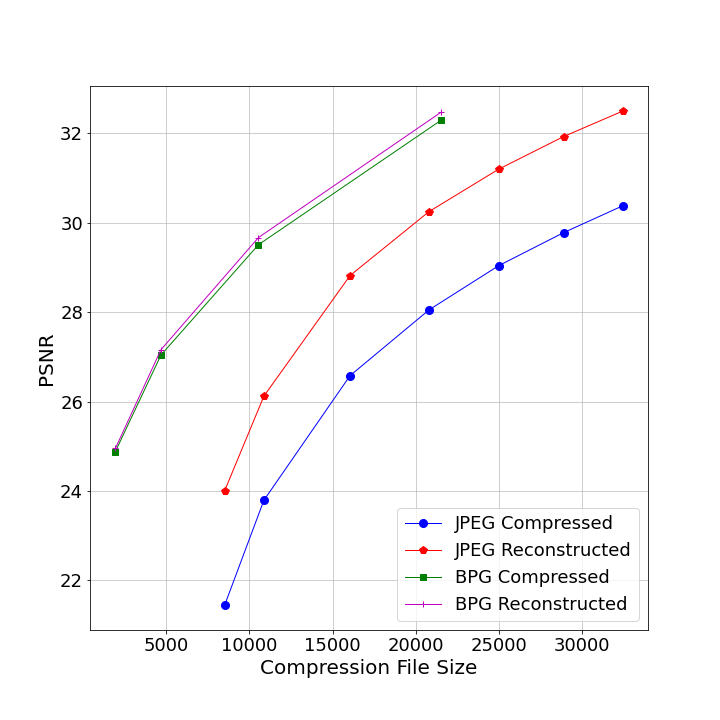}}
\subfloat[][PSNR with down-scaling operation]{\includegraphics[scale=0.34]{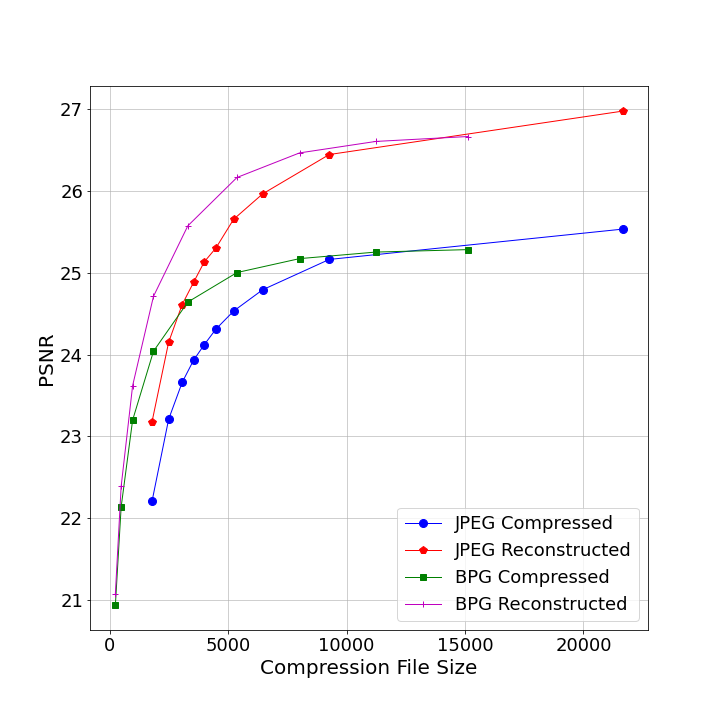}}
\caption{Variation of the average of SSIM (a),PSNR (b) versus image quality for Kodak test images without or with down-scaling.}%
\label{fig:SSIMPSNRversusQuality}%
\end{figure*}

\section{Experimental Analysis}
\label{sec:performance}
In this section, we include the implementation details of the proposed solution. Then, we present and discuss the evaluation results including the compression visual effect, the communication size efficiency, the visual quality enhancement after applying the proposed model and the power consumption. Experiments were done on a Tesla V100 GPU. 

\subsection{Data Description}
The RDN model was trained  using the dataset described in~\cite{mentzer2020high}. This dataset consists of a large set of colored images collected from the Internet. These high-resolution images are first down-scaled to images having random size between 500 and 1000 pixels. Then, the obtained images of variable size are cropped randomly to get 256x256 images. These images are compressed using JPEG and BPG, respectively, and the obtained pixel values are scaled between [0;1]. The testing is done with images chosen from the Kodak dataset (24 images)~\cite{www:r0k.us}. 

\subsection{Model Implementation}
 The employed RDN model was implemented in Pytorch~\cite{RRDBgithub}. The mini-batch technique is used to train the RDN model. The learning rate is initialized to $10^{-4}$. At each iteration, normalization is applied to each mini-batch. The final output (desired output) is the reference uncompressed image (enhanced) $I$, which will be compared with the compressed image through the loss function. The Adam optimizer with an adaptive learning factor is used to optimize the loss function. \\

Before being passed to the RDN model, the images are compressed by JPEG and BPG, respectively, with variable compression ratio. Increasing the compression ratio will lead to an increase in the block artifacts. The size of the image patch should be then selected to include relevant useful patterns. According to the empirical evaluation, we find that the size of the image patch varies between 96 and 192 to contain enough information to remove noise and compression artifacts.\\ 

Two models are trained, where the first model is for the first variant (compression image without down-scaling operation), and the second one is for the second variant (compression with down-scaling operation). The compressed (and scaled) and uncompressed images are then fed into the proposed model. The output of the first RDN model (first variant) has the same size as the input, which can be considered as the input image plus the related compressed noised image. While for the second model of the second variant, the input size is down-scaled by 4 compared to the output high-resolution image, which has the same size as the original uncompressed one. These RDN models can be then employed to better remove and/or reduce the compressed noise.

\subsection{Compression Effect on Image Quality}
In this part, we analyze the effect of the compression on the visual degradation, considering different compression ratios. The lower is the compression ratio, the better is the compressed image quality, and vice-versa. The image quality parameter depends on the target IoT multimedia application and should be set accordingly (see \figurename~\ref{fig:SSIMversusQualityCmpression}). We should note that this parameter has not to be static and could be adapted dynamically regarding the instantaneous requirement of the IoT multimedia application. For instance, if an event of major importance occurs (e.g. robber detection in a surveillance system) and the application requires a higher image quality, it could then configure the MIoT devices to send the image with low compression ratio. Otherwise, the application server (or data center) could enhance the quality of highly compressed images. In this sense, the proposed approach can offer the flexibility on the quality of the received multimedia contents that can be controlled by the application server (or the data center). \\

\begin{figure}[!ht]
\centering
\subfloat[][kodim01]{\includegraphics[scale=0.125]{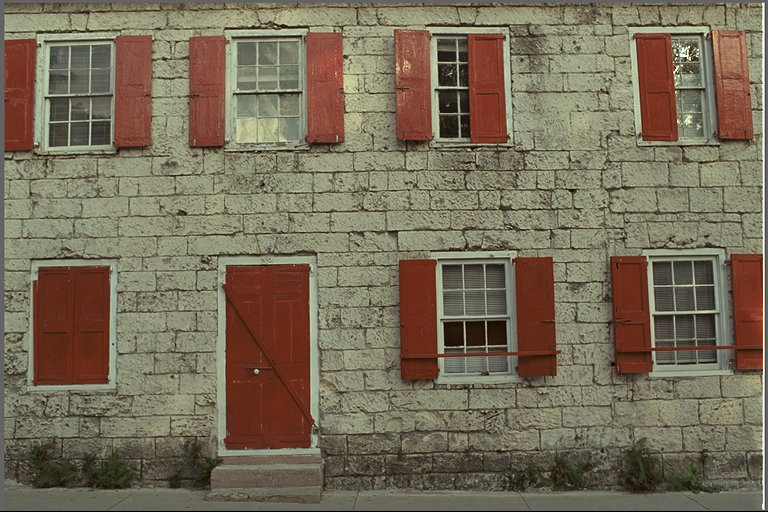}}\hfill
\subfloat[][kodim02]{\includegraphics[scale=0.125]{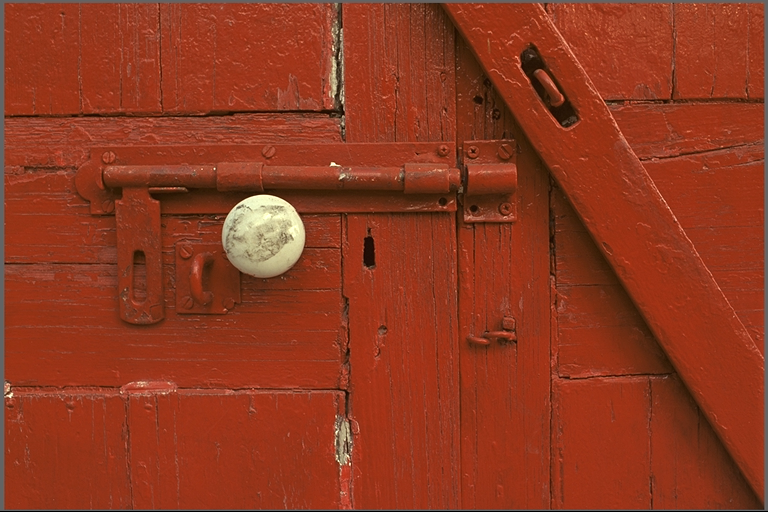}}\hfill
\caption{Original uncompressed images "kodim01",(a) and "kodim02" (b)}
\label{fig:orig}%
\end{figure}

The variation of Structural Similarity (SSIM) and Peak Signal-to-Noise Ratio (PSNR) versus compressed data size are presented for both variants in \figurename~\ref{fig:SSIMPSNRversusQuality} for the images shown in \figurename~\ref{fig:orig}. The obtained results indicate that by increasing the image quality (or decreasing the compression ratio), the PSNR, SSIM, and the size of the transmitted (stored) data increase. This clearly indicates that the visual degradation increases when the compressed data size decreases (or when the compression ratio increases), as shown in \figurename~\ref{fig:SSIMPSNRversusQuality}. Thus, a trade-off between visual quality and compressed data size exists. 

\begin{figure*}[!ht]
\centering
\subfloat[][Quality=1]{\includegraphics[scale=0.125]{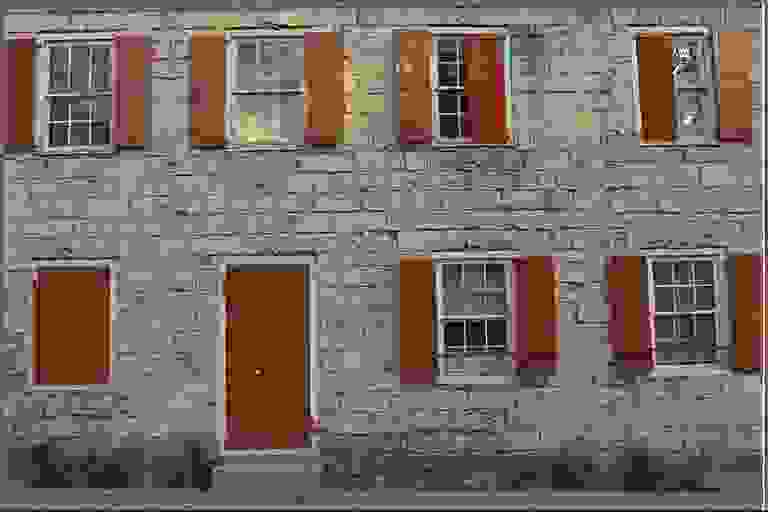}}\hfill
\subfloat[][Quality=5]{\includegraphics[scale=0.125]{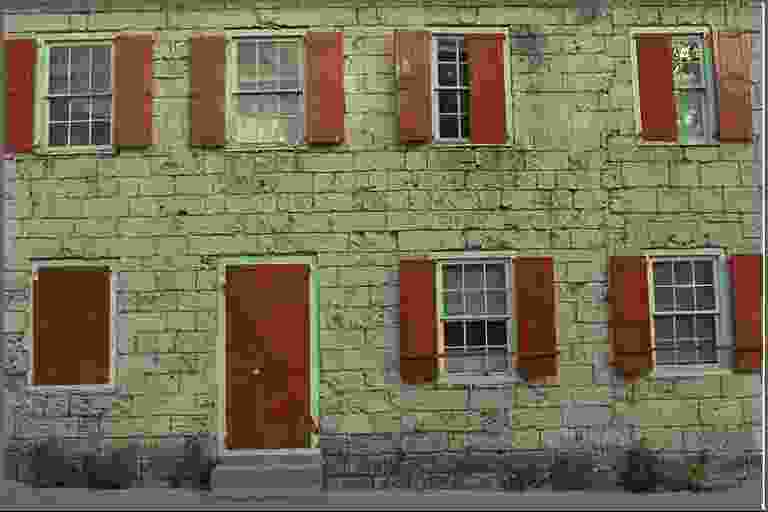}}\hfill
\subfloat[][Quality=10]{\includegraphics[scale=0.125]{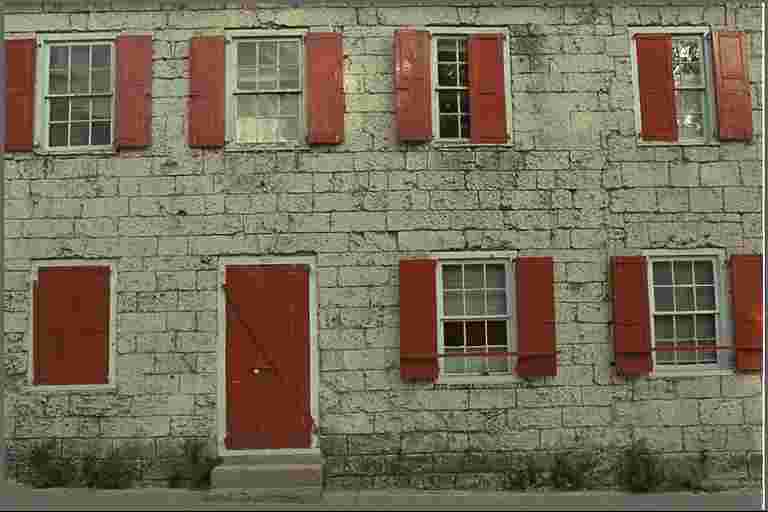}}\hfill
\subfloat[][Quality=15]{\includegraphics[scale=0.125]{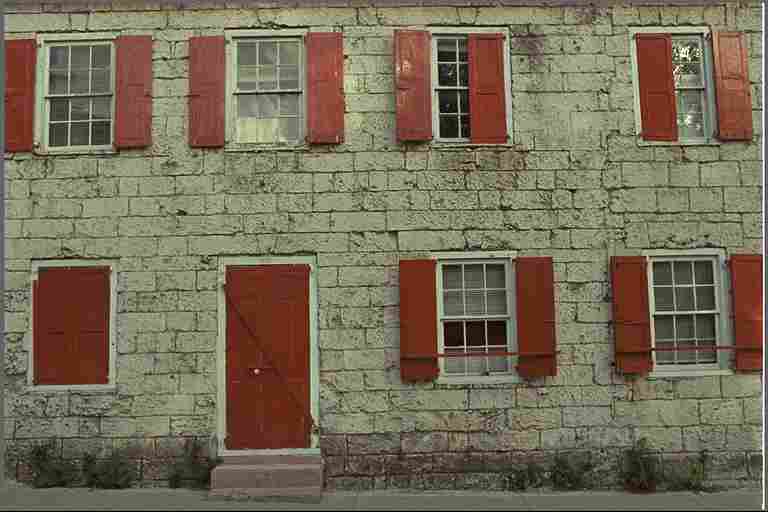}}\hfill
\\
\subfloat[][Quality=1]{\includegraphics[scale=0.125]{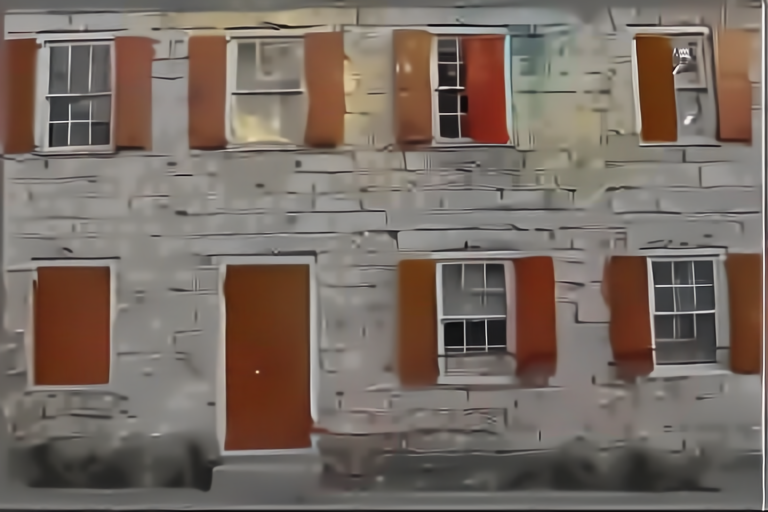}}\hfill
\subfloat[][Quality=5]{\includegraphics[scale=0.125]{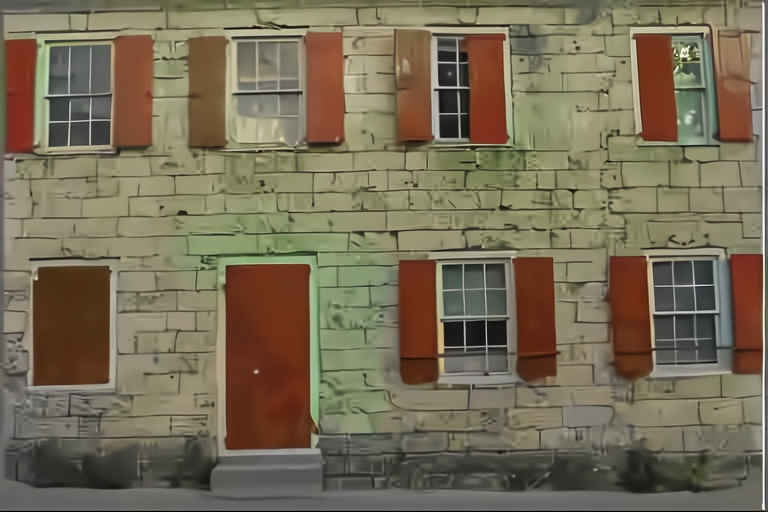}}\hfill
\subfloat[][Quality=10]{\includegraphics[scale=0.125]{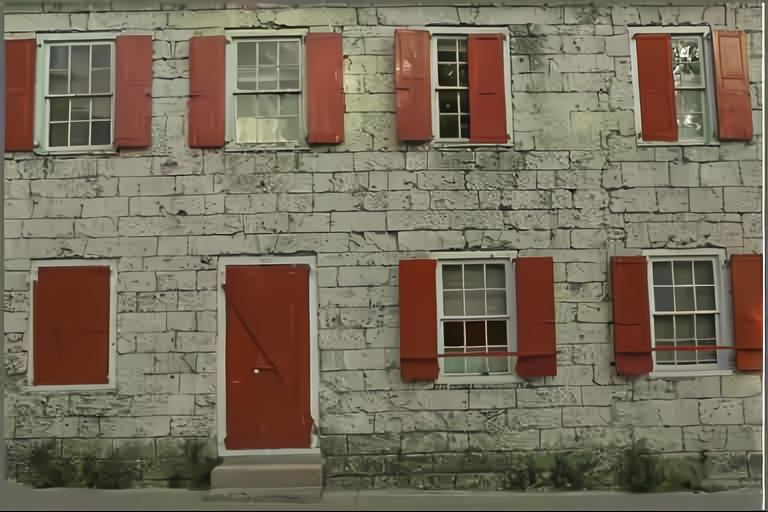}}\hfill
\subfloat[][Quality=15]{\includegraphics[scale=0.125]{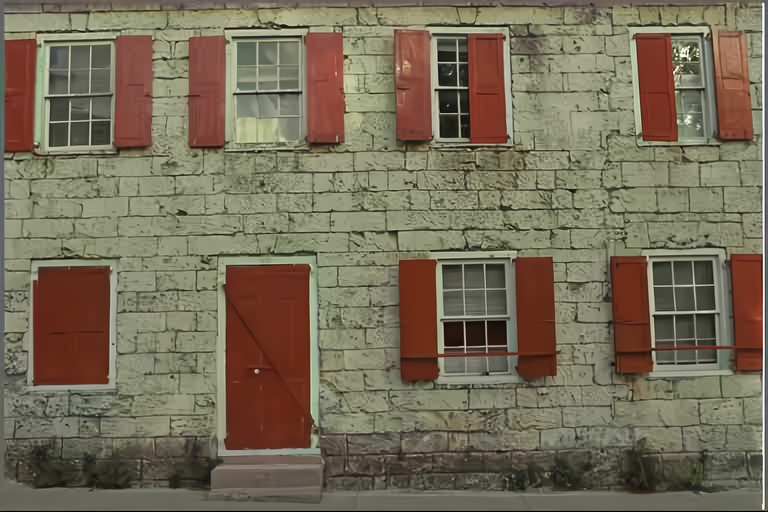}}\hfill

\caption{Corresponding compressed (JPEG) "kodim01" images (a)-(e) and reconstructed ones (f)-(j) by using the proposed model after decompression, respectively versus different image quality, and for the proposed solution without down-scaling operation.}%
\label{fig:CompressEnhanced}%
\end{figure*}

\begin{figure*}[!ht]
\centering
\subfloat[][Quality=1]{\includegraphics[scale=0.125]{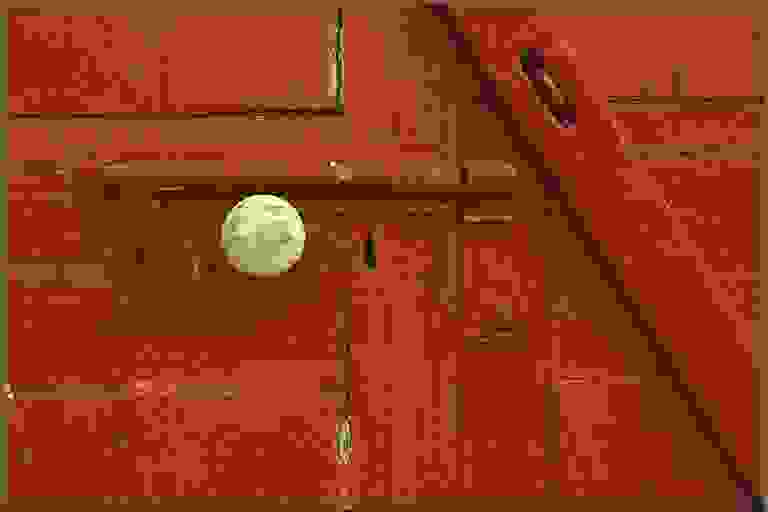}}\hfill
\subfloat[][Quality=5]{\includegraphics[scale=0.125]{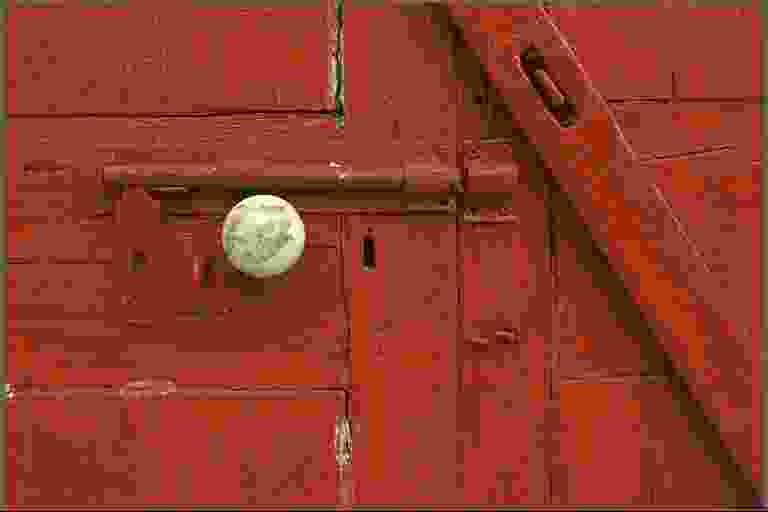}}\hfill
\subfloat[][Quality=10]{\includegraphics[scale=0.125]{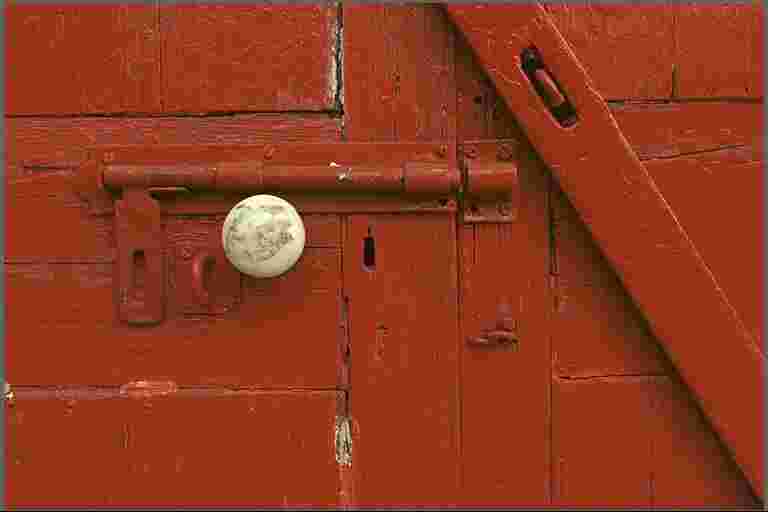}}\hfill
\subfloat[][Quality=15]{\includegraphics[scale=0.125]{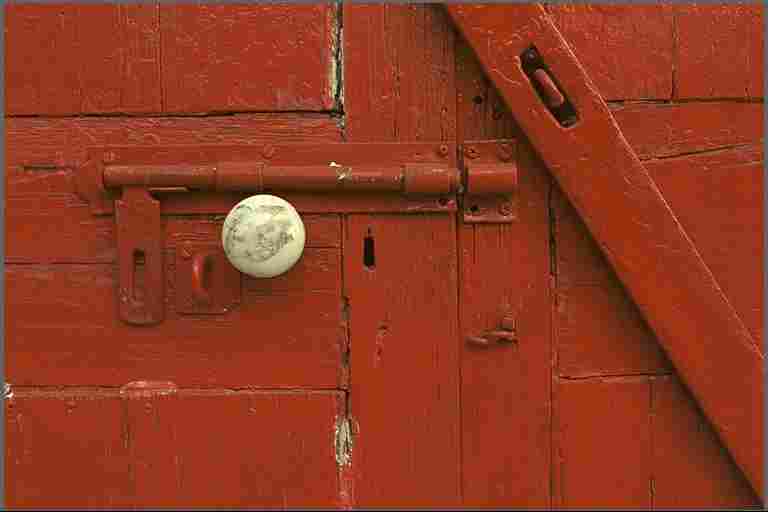}}\hfill\\
\subfloat[][Quality=1]{\includegraphics[scale=0.125]{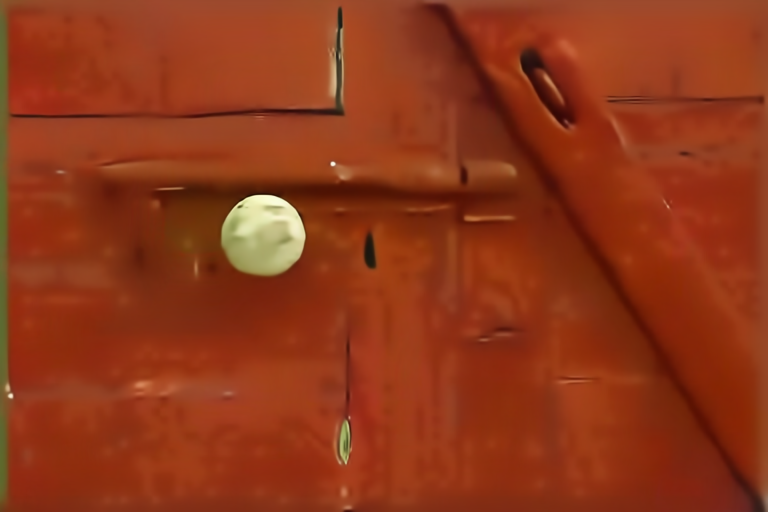}}\hfill
\subfloat[][Quality=5]{\includegraphics[scale=0.125]{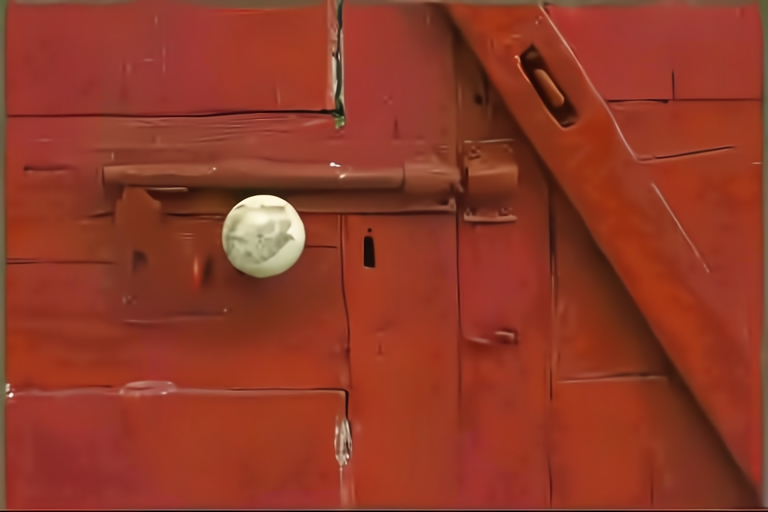}}\hfill
\subfloat[][Quality=10]{\includegraphics[scale=0.125]{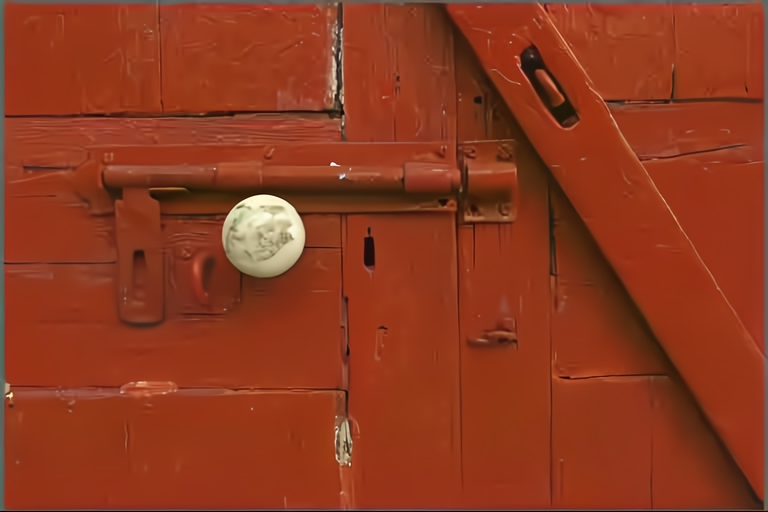}}\hfill
\subfloat[][Quality=15]{\includegraphics[scale=0.125]{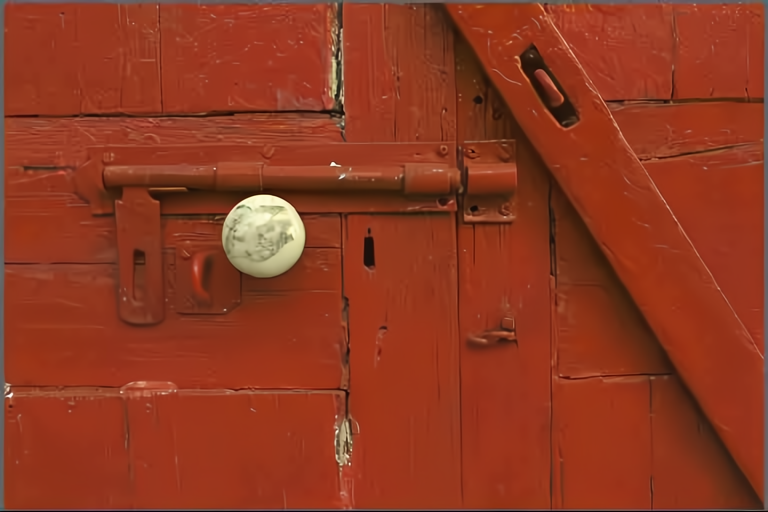}}\hfill

\caption{Corresponding compressed (JPEG) "kodim02" images (a)-(e) and reconstructed ones (f)-(j) by using the proposed model after decompression, respectively versus different image quality, and for the proposed solution without down-scaling operation.}%
\label{fig:CompressEnhanced2}%
\end{figure*}

\begin{figure*}[!ht]
\centering
\subfloat[][Quality=10]{\includegraphics[scale=0.125]{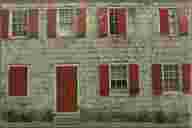}}\hfill
\subfloat[][Quality=20]{\includegraphics[scale=0.125]{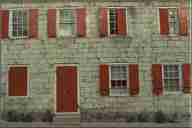}}\hfill
\subfloat[][Quality=30]{\includegraphics[scale=0.125]{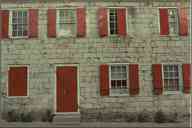}}\hfill
\subfloat[][Quality=40]{\includegraphics[scale=0.125]{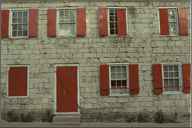}}\hfill\\
\subfloat[][Quality=10]{\includegraphics[scale=0.125]{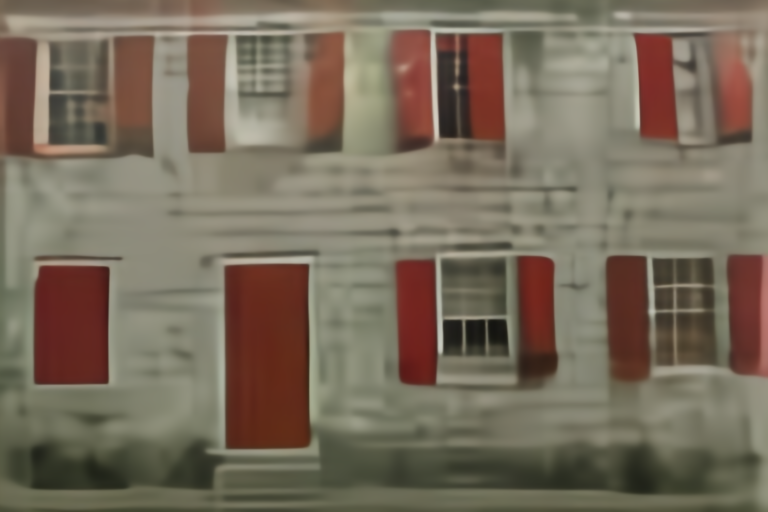}}\hfill
\subfloat[][Quality=20]{\includegraphics[scale=0.125]{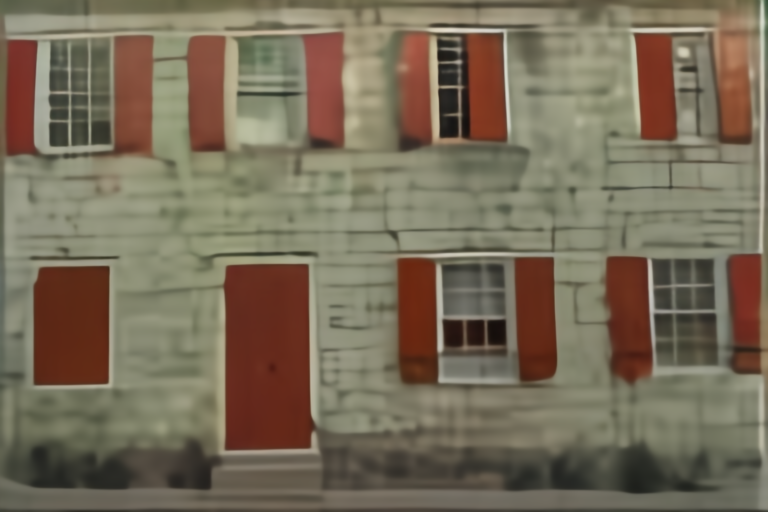}}\hfill
\subfloat[][Quality=30]{\includegraphics[scale=0.125]{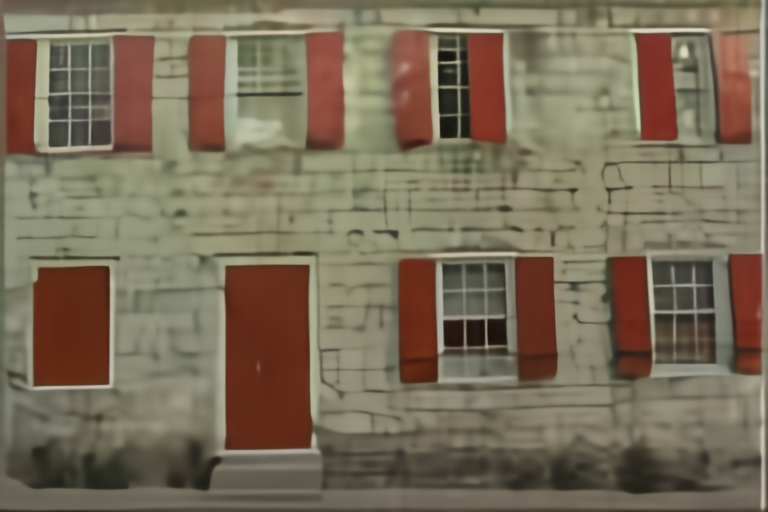}}\hfill
\subfloat[][Quality=40]{\includegraphics[scale=0.125]{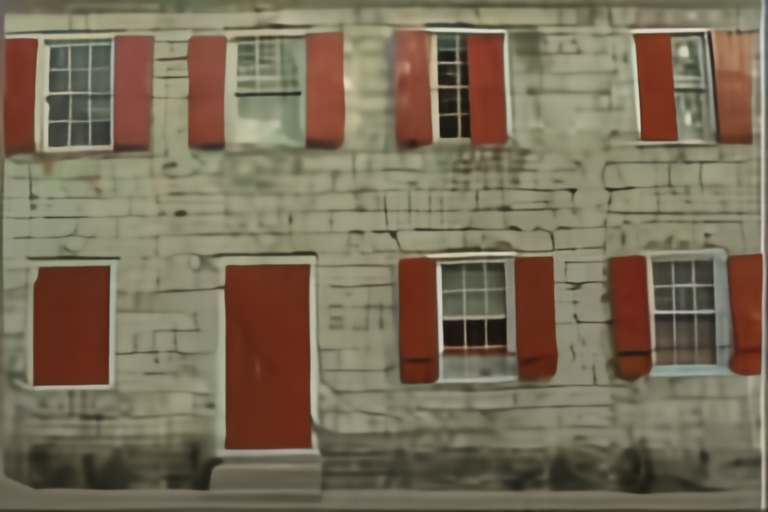}}\hfill
\caption{Corresponding compressed scaled (JPEG) "kodim01" images (a)-(e) and the reconstructed ones (f)-(j) by using the proposed model after decompression, respectively versus different image quality, and for the proposed solution with down-scaling operation.}%
\label{fig:CompressEnhancedscale}%
\end{figure*}

\begin{figure*}[!ht]
\centering
\subfloat[][Quality=10]{\includegraphics[scale=0.125]{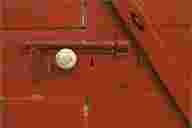}}\hfill
\subfloat[][Quality=20]{\includegraphics[scale=0.125]{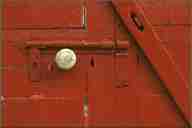}}\hfill
\subfloat[][Quality=30]{\includegraphics[scale=0.125]{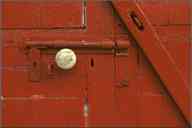}}\hfill
\subfloat[][Quality=40]{\includegraphics[scale=0.125]{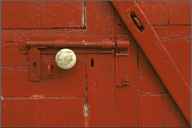}}\hfill
\\

\subfloat[][Quality=10]{\includegraphics[scale=0.125]{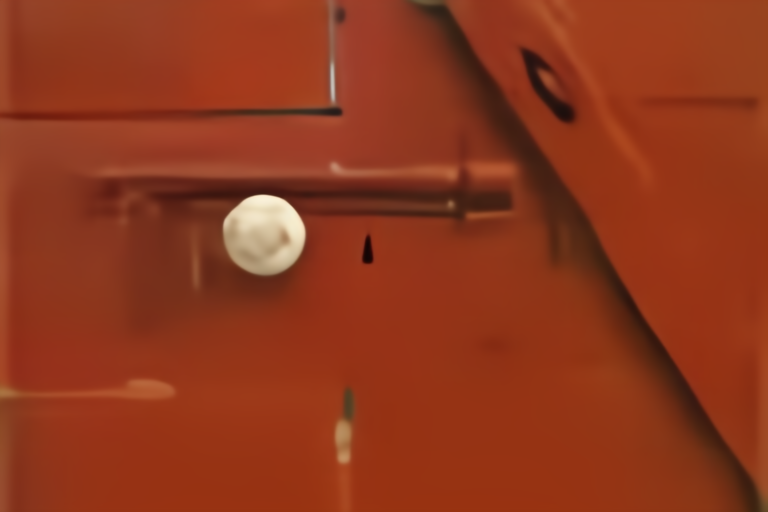}}\hfill
\subfloat[][Quality=20]{\includegraphics[scale=0.125]{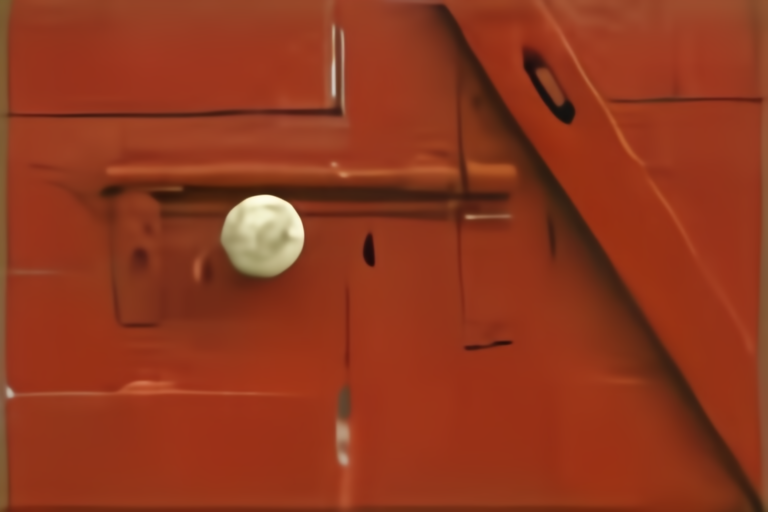}}\hfill
\subfloat[][Quality=30]{\includegraphics[scale=0.125]{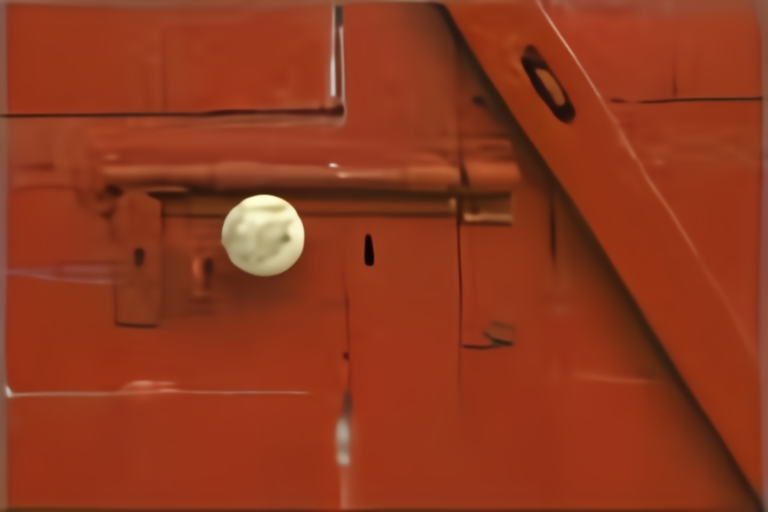}}\hfill
\subfloat[][Quality=40]{\includegraphics[scale=0.125]{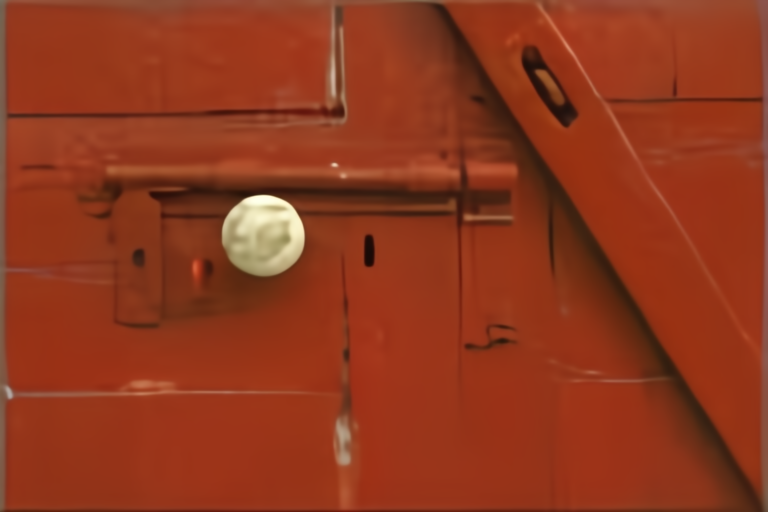}}\hfill
\caption{Corresponding compressed scaled (JPEG) "kodim02" images (a)-(e) and the reconstructed ones (f)-(j) by using the proposed model after decompression, respectively versus different image quality, and for the proposed solution with down-scaling operation.}%
\label{fig:CompressEnhancedscale2}%
\end{figure*}

\subsection{Denoising and Super-resolution Effect on Image Quality}

Given that some visual information are lost after lossy compression and cannot be recovered, the proposed approach is introduced to provide a better image quality by reducing the compression noise effect. In this context, the results shown in \figurename~\ref{fig:SSIMPSNRversusQuality} show that the SSIM and PSNR values of the enhanced reconstructed images are higher compared to the decompressed ones with or without down-scaling operations. Moreover, the obtained visual content quality with the first variant is better compared to the second variant. Therefore, if the down-scaling operation is applied (second variant), a lower compression ratio is required compared to the first variant towards preserving the main visual content quality. However, the down-scaling results in higher data reduction compared to the first variant (without down-scaling) even with low compression ratio. Thus, the choice of which variant should be applied (with or without down-scaling operation) depends on the target IoT multimedia application and MIoT device limitations. For example, 33\% of compressed data reduction compared to a JPEG image with default image quality (85\%) is required to reach the same image quality by using the proposed approach (with the first variant). \\

To evaluate the image quality enhancement, the proposed solution (two different models for each image compression algorithm) was tested with 24 images selected from the kodak dataset. For two images randomly chosen from the testing dataset (see \figurename~\ref{fig:orig}), we show the visual degradation versus image compression ratio for the decompressed and enhanced images with the first variant (without down-scaling operation) in \figurename~\ref{fig:CompressEnhanced} and~\ref{fig:CompressEnhanced2}, respectively. Also, same results are provided with the second variant (with down-scaling operation) in \figurename~\ref{fig:CompressEnhancedscale} and~\ref{fig:CompressEnhancedscale2}. It should be noted here that these results are for the JPEG standard, and that similar results were obtained with BPG, but they were omitted to conserve space.\\


To evaluate the visual degradation before and after applying the proposed model, the PSNR and SSIM metrics were used. According to the obtained results, illustrated in \figurename~\ref{fig:SSIMPSNRversusQuality}, the proposed solution ensures better image quality compared to the decompressed one. This indicates clearly that the effect of lossy compression with high compression ratio can be reduced by applying the proposed RDN model (without or with down-scaling). Moreover, the average of SSIM and PSNR, shown in \figurename~\ref{fig:SSIMPSNRversusQuality}, indicates that the recovered images with the first variant presents higher SSIM and PSNR values compared to the second variant that uses a down-scaling operation.  

In the JPEG case (see \figurename~\ref{fig:SSIMPSNRversusQuality}-a)), with the first variant, the SSIM varies between 0.5532 and 0.8574, while for the reconstructed images, the SSIM is between 0.6477 and 0.8936. Similarly, the PSNR varies between 21.44 dB and 30.38 dB for the decompressed images, and it varies between 24dB and 32.5dB for the reconstructed images. For the first variant, the enhancement in the visual content is increased by 0.0585 in terms of SSIM and by 2.25 in terms of PSNR. Additionally, we found that on average, if the compressed file size is increased by 3989 bytes, the SSIM and PSNR values will be increased by 0.041 and 1.417, respectively. Besides, with the second variant, the SSIM varies between 0.5415 and 0.6809 for the decompressed images, while the SSIM of the enhanced recovered images varies between 0.5881 and 0.7415. The enhancement of the visual content quality is increased by 0.0491 in terms of SSIM and by 1.09 in terms of PSNR. Additionally, we found that on average, if the compressed file size is increased by 574 bytes, the SSIM and PSNR values will be increased by 0.015 and 0.387, respectively. Moreover, for the first variant, the maximum SSIM and PSNR values are 0.89 and 32 dB, respectively, while for the second variant, the maximum values of SSIM and PSNR are 0.741 and 26, respectively. Furthermore, the second variant with an average of 18\% compression size of the first variant can ensure less than 0.168 for the SSIM and 4.56 for the PSNR. The first variant with JPEG ensures an SSIM and PSNR greater by an average of 0.15 and 3.3 in the first variant compared to the second one, respectively but with 4 times of additional communication overhead. \\

 In the case of BPG (see \figurename~\ref{fig:SSIMPSNRversusQuality}-b)), a lower visual content enhancement is achieved for the first variant compared to JPEG. This is due to the fact that decompressed BPG images are less degraded compared to the JPEG case. Furthermore, BPG introduces more computing overhead (more resources for the compression process compared to JPEG) but it ensures less communication overhead. In contrast, for the second variant, the visual content is enhanced with BPG, where a down-scaling operation is applied. The SSIM varies between 0.6355 and 0.8804 for the reconstructed enhanced images with the first variant, while it varies between 0.5374 and 0.7394 for the second variant. Similarly, the PSNR varies between 24.95 dB and 32.48 dB for the first variant, and it varies between 21.08 dB and 26.66 dB for the second variant. The enhancement in the visual content is increased on average by 0.00375 and 0.0373 in terms of SSIM and by 0.1325 and 0.8433 in terms of PSNR for the first and second variants, respectively.


\begin{table*}[!ht]
\caption{Comparative analysis results between proposed variants}
\begin{tabular}{|p{3cm}|p{3cm}|p{3cm}|p{3cm}|p{3cm}|}
\hline
\textbf{Image Quality without Down-scaling} & \textbf{Image Quality with Down-scaling} & \textbf{SSIM without down-scaling-SSIM with down-scaling)} & \textbf{PSNR without down-scaling-PSNR with down-scaling)} & \textbf{Compression Size Ratio (without down-scaling/with down-scaling)} \\ \hline
         1  & 10 & 0.012  &-0.77  & 0.209                              \\ \hline
         5  & 20 & 0.07   & 0.58  & 0.228                             \\ \hline
         10 & 30& 0.152  & 2.92  &  0.19                             \\ \hline
         15 & 40 & 0.191  & 4.12  & 0.17                              \\ \hline
         20 & 50 & 0.213  & 4.92  & 0.16                              \\ \hline
         25 & 60 & 0.225  & 5.47  & 0.155                            \\ \hline
         30 & 70& 0.23   &  5.85 & 0.161                            \\ \hline
\end{tabular}
\label{tab:imageQuality}
\end{table*}





\subsection{Communication Size Reduction}
The communication efficiency is proportional to the communicated data size. In our proposed approach, we aim at adapting the existing compression algorithms to be applied with high compression ratio to achieve communication efficiency. The compression ratio is calculated by dividing the compressed image size by the original image size. Consequently, a higher compression ratio can be obtained with smaller compressed data size but this leads to an important visual degradation in the decompressed image. In the proposed model, a higher compression ratio can be employed while achieving same image quality as in the standard case with lower compression ratio. For example, in the case of JPEG, lower image quality can be used  (between 1 and 25\% for the first variant, and between 20\% and 60\% for the second variant). For different image quality (compression ratio) values, the corresponding ratio between the compressed transmitted data with scaling and without scaling are presented in \tablename~\ref{tab:imageQuality}. This clearly indicates that the proposed solution with down-scaling reduces the size of the compressed image by 18\% on average compared to the first variant.

\subsection{Transmission Power Consumption}
In resources-constrained MIoT devices, an efficient multimedia compression algorithm should reduce the communication overhead and consequently the transmission power consumption without compromising the image quality. The advantage of the proposed solution is that it can reduce more the transmitted data size (using higher compression ratio) compared to existing compression configurations while preserving the image quality since a super-resolution denoising DL model is applied at the application server. As shown in \figurename~\ref{fig:SSIMPSNRversusQuality}, the obtained results (PSNR and SSIM) show that an acceptable visual quality is reached with very low compression size (for both variants and image compression algorithms). Moreover, the down-scaling operation in the second variant reduces more the communication overhead compared to the first variant. To obtain acceptable image quality, on average 18\% of the compressed image size is required compared to the first variant for a down-sampling factor equals to 4. Thus, the second variant ensures a significant energy consumption reduction and consequently 80\% of MIoT device transmission energy is preserved but with an additional visual degradation compared to the first variant. The proposed solution decreases by approximately 4 times the transmission power of the MIoT devices (in case the second variant is employed).

\section{Discussions}
\label{sec:discussion}
In this paper, the aim was to minimize the communication cost by highly compressing the data without loosing the image quality. The work done in this paper can improve the existing limitations of multimedia compression approaches used in MIoT. The advantage of the proposed solution is that it is independent of the used multimedia compression algorithms and it does not require any modification in the existing multimedia compression standards (practical implementation). The main challenge was to preserve image quality while using high compression ratio.  This was addressed by employing the RDN learning model. The proposed solution was tested with the JPEG and BPG image compression algorithms (with or without down-scaling operation).\\

The employed RDN learning model succeeded in enhancing the compressed image quality. This was validated by evaluating the PSNR and SSIM values that were increased compared to the decompressed images for both variants. This solution was designed to not require any additional operation on the MIoT devices by delegating the role of image quality enhancement to the application server. In addition, the proposed solution optimizes the transmitted data size, that can be reduced to 33\% for image quality equal to 5 with JPEG compression compared to default image quality. Thus, the proposed solution could highly improve the transmission efficiency for any MIoT application using any compressor like JPEG or BPG. The experimentation results showed that the average PSNR and SSIM values are increased and the transmission data ratio is decreased by an average of 18\% using the second variant compared to the first variant. \\

 A further optimization of the communicated data size would be in delivering gray images while designing (using) a DL-based model for the colorization. As a future work, we will further explore other learning models that can colorize gray images towards reducing more and more the communication overhead.

\section{Conclusion}
\label{sec:conclusion}
In this paper, we aim to respond better to the hard challenge of communication and resources overhead in the MIoT domain. One solution to cope with the high communicated data size is to use hard lossy multimedia data compression. In this context, a high compression ratio is required to achieve high communication efficiency. However, the current filtering theory cannot fix the hard effect of multimedia compression with a high compression ratio. Our proposed solution consists of reducing the hard visual degradation at the application server by employing a denoising-super resolution DL-based model. Two variants of the proposed scheme are considered: the first one applies compression with a high image compression ratio but without down-scaling the image size, while the second one down-scales the image before applying the lossy compression. The second variant ensures minimum communication overhead at the cost of additional visual degradation compared to the first variant. The experimentation results showed that enhanced decompressed images are obtained (with or without down-scaling) being compressed with high compression ratio.  Equally important, the performance analysis confirms the effectiveness of the proposed solution since it reaches a good balance between visual degradation and communication size. Thus, the proposed solution can be considered as an adequate candidate to enhance compressed MIoT transmitted data. Finally, one of the main features of the proposed approach is being flexible, and functional with any multimedia compressor at different compression ratios and down-scaling factor.

\section*{Acknowledgment}
This work was  partially  supported  by  the  EIPHI  Graduate  School(contract ”ANR-17-EURE-0002”). The Mesocentre of Franche-Comté provided the computing facilities.

\bibliographystyle{unsrt}

\begin{thebibliography}{10}

\bibitem{goelreview}
Shalini~Sharma Goel, Anubhav Goel, Mohit Kumar, and Germ{\'a}n Molt{\'o}.
\newblock A review of internet of things: qualifying technologies and boundless
  horizon.
\newblock {\em Journal of Reliable Intelligent Environments}, pages 1--11,
  2021.

\bibitem{wang2017multimedia}
Qin Wang, Yanxiao Zhao, Wei Wang, Daniel Minoli, Kazem Sohraby, Hongbo Zhu, and
  Ben Occhiogrosso.
\newblock Multimedia iot systems and applications.
\newblock In {\em 2017 Global Internet of Things Summit (GIoTS)}, pages 1--6.
  IEEE, 2017.
  
\bibitem{yaacoub2020securing}
Jean-Paul A Yaacoub, Mohamad Noura, Hassan N Noura, Ola Salman, Elias Yaacoub, Rapha{\"e}l Couturier, and Ali Chehab.
 \newblock Securing internet of medical things systems: limitations, issues and recommendations.
\newblock {\em Future Generation Computer Systems},105:581--606, 2020.


\bibitem{tanwar2019multimedia}
Sudeep Tanwar, Sudhanshu Tyagi, and Neeraj Kumar.
\newblock {\em Multimedia big data computing for IoT applications: Concepts,
  paradigms and solutions}, volume 163.
\newblock Springer, 2019.

\bibitem{nauman2020multimedia}
Ali Nauman, Yazdan~Ahmad Qadri, Muhammad Amjad, Yousaf~Bin Zikria,
  Muhammad~Khalil Afzal, and Sung~Won Kim.
\newblock Multimedia internet of things: A comprehensive survey.
\newblock {\em IEEE Access}, 8:8202--8250, 2020.

\bibitem{al2020survey}
Mohammed~Ali Al-Garadi, Amr Mohamed, Abdulla Al-Ali, Xiaojiang Du, Ihsan Ali,
  and Mohsen Guizani.
\newblock A survey of machine and deep learning methods for internet of things
  (iot) security.
\newblock {\em IEEE Communications Surveys \& Tutorials}, 2020.

\bibitem{jan2017balanced}
Naeem Jan, Nadeem Javaid, Qaisar Javaid, Nabil Alrajeh, Masoom Alam, Zahoor~Ali
  Khan, and Iftikhar~Azim Niaz.
\newblock A balanced energy-consuming and hole-alleviating algorithm for
  wireless sensor networks.
\newblock {\em IEEE Access}, 5:6134--6150, 2017.

\bibitem{li2018learning}
He~Li, Kaoru Ota, and Mianxiong Dong.
\newblock Learning iot in edge: Deep learning for the internet of things with
  edge computing.
\newblock {\em IEEE network}, 32(1):96--101, 2018.

\bibitem{tang2017enabling}
Jie Tang, Dawei Sun, Shaoshan Liu, and Jean-Luc Gaudiot.
\newblock Enabling deep learning on iot devices.
\newblock {\em Computer}, 50(10):92--96, 2017.

\bibitem{tian2018deep}
Chunwei Tian, Yong Xu, Lunke Fei, and Ke~Yan.
\newblock Deep learning for image denoising: a survey.
\newblock In {\em International Conference on Genetic and Evolutionary
  Computing}, pages 563--572. Springer, 2018.

\bibitem{tian2020deep}
Chunwei Tian, Lunke Fei, Wenxian Zheng, Yong Xu, Wangmeng Zuo, and Chia-Wen
  Lin.
\newblock Deep learning on image denoising: An overview.
\newblock {\em Neural Networks}, 2020.

\bibitem{bai2020survey}
K~Bai, X~Liao, Q~Zhang, X~Jia, and S~Liu.
\newblock Survey of learning based single image super-resolution reconstruction
  technology.
\newblock {\em Pattern Recognition and Image Analysis}, 30(4):567--577, 2020.

\bibitem{wang2020deep}
Zhihao Wang, Jian Chen, and Steven~CH Hoi.
\newblock Deep learning for image super-resolution: A survey.
\newblock {\em IEEE transactions on pattern analysis and machine intelligence},
  2020.

\bibitem{zhang2018residual}
Yulun Zhang, Yapeng Tian, Yu~Kong, Bineng Zhong, and Yun Fu.
\newblock Residual dense network for image super-resolution.
\newblock In {\em Proceedings of the IEEE conference on computer vision and
  pattern recognition}, pages 2472--2481, 2018.

\bibitem{wallace1992jpeg}
Gregory~K Wallace.
\newblock The jpeg still picture compression standard.
\newblock {\em IEEE transactions on consumer electronics}, 38(1):xviii--xxxiv,
  1992.

\bibitem{skodras2001jpeg}
Athanassios Skodras, Charilaos Christopoulos, and Touradj Ebrahimi.
\newblock The jpeg 2000 still image compression standard.
\newblock {\em IEEE Signal processing magazine}, 18(5):36--58, 2001.

\bibitem{BPGIm8857370:online}
F.~Bellard.
\newblock Bpg image format.
\newblock \url{https://bellard.org/bpg/}, 12 2014.

\bibitem{albalawi2015hardware}
Umar Albalawi, Saraju~P Mohanty, and Elias Kougianos.
\newblock A hardware architecture for better portable graphics (bpg)
  compression encoder.
\newblock In {\em 2015 IEEE International Symposium on Nanoelectronic and
  Information Systems}, pages 291--296. IEEE, 2015.

\bibitem{sze2014high}
Vivienne Sze, Madhukar Budagavi, and Gary~J Sullivan.
\newblock High efficiency video coding (hevc).
\newblock In {\em Integrated circuit and systems, algorithms and
  architectures}, volume~39, page~40. Springer, 2014.

\bibitem{wiegand2003overview}
Thomas Wiegand, Gary~J Sullivan, Gisle Bjontegaard, and Ajay Luthra.
\newblock Overview of the h. 264/avc video coding standard.
\newblock {\em IEEE Transactions on circuits and systems for video technology},
  13(7):560--576, 2003.

\bibitem{dong2014learning}
Chao Dong, Chen~Change Loy, Kaiming He, and Xiaoou Tang.
\newblock Learning a deep convolutional network for image super-resolution.
\newblock In {\em European conference on computer vision}, pages 184--199.
  Springer, 2014.

\bibitem{dong2016accelerating}
Chao Dong, Chen~Change Loy, and Xiaoou Tang.
\newblock Accelerating the super-resolution convolutional neural network.
\newblock In {\em European conference on computer vision}, pages 391--407.
  Springer, 2016.

\bibitem{kim2016accurate}
Jiwon Kim, Jung Kwon~Lee, and Kyoung Mu~Lee.
\newblock Accurate image super-resolution using very deep convolutional
  networks.
\newblock In {\em Proceedings of the IEEE conference on computer vision and
  pattern recognition}, pages 1646--1654, 2016.

\bibitem{kim2016deeply}
Jiwon Kim, Jung Kwon~Lee, and Kyoung Mu~Lee.
\newblock Deeply-recursive convolutional network for image super-resolution.
\newblock In {\em Proceedings of the IEEE conference on computer vision and
  pattern recognition}, pages 1637--1645, 2016.

\bibitem{tai2017image}
Ying Tai, Jian Yang, and Xiaoming Liu.
\newblock Image super-resolution via deep recursive residual network.
\newblock In {\em Proceedings of the IEEE conference on computer vision and
  pattern recognition}, pages 3147--3155, 2017.

\bibitem{tai2017memnet}
Ying Tai, Jian Yang, Xiaoming Liu, and Chunyan Xu.
\newblock Memnet: A persistent memory network for image restoration.
\newblock In {\em Proceedings of the IEEE international conference on computer
  vision}, pages 4539--4547, 2017.

\bibitem{ledig2017photo}
Christian Ledig, Lucas Theis, Ferenc Husz{\'a}r, Jose Caballero, Andrew
  Cunningham, Alejandro Acosta, Andrew Aitken, Alykhan Tejani, Johannes Totz,
  Zehan Wang, et~al.
\newblock Photo-realistic single image super-resolution using a generative
  adversarial network.
\newblock In {\em Proceedings of the IEEE conference on computer vision and
  pattern recognition}, pages 4681--4690, 2017.

\bibitem{tong2017image}
Tong Tong, Gen Li, Xiejie Liu, and Qinquan Gao.
\newblock Image super-resolution using dense skip connections.
\newblock In {\em Proceedings of the IEEE International Conference on Computer
  Vision}, pages 4799--4807, 2017.

\bibitem{haris2018deep}
Muhammad Haris, Gregory Shakhnarovich, and Norimichi Ukita.
\newblock Deep back-projection networks for super-resolution.
\newblock In {\em Proceedings of the IEEE conference on computer vision and
  pattern recognition}, pages 1664--1673, 2018.

\bibitem{wang2018deep}
Yucheng Wang, Jialiang Shen, and Jian Zhang.
\newblock Deep bi-dense networks for image super-resolution.
\newblock In {\em 2018 Digital Image Computing: Techniques and Applications
  (DICTA)}, pages 1--8. IEEE, 2018.

\bibitem{li2021rgsr}
Biao Li, Yong Shi, Bo~Wang, Zhiquan Qi, and Jiabin Liu.
\newblock Rgsr: A two-step lossy jpg image super-resolution based on noise
  reduction.
\newblock {\em Neurocomputing}, 419:322--334, 2021.

\bibitem{mentzer2020high}
Fabian Mentzer, George~D Toderici, Michael Tschannen, and Eirikur Agustsson.
\newblock High-fidelity generative image compression.
\newblock {\em Advances in Neural Information Processing Systems}, 33, 2020.

\bibitem{www:r0k.us}
True color kodak images.
\newblock \url{http://r0k.us/graphics/kodak/}, 06 2004.

\bibitem{RRDBgithub}
cszn/kair: Image restoration toolbox pytorch training and testing codes for
  usrnet dncnn ffdnet srmd dpsr esrgan.
\newblock \url{https://github.com/cszn/KAIR}, 10 2020.

\end{thebibliography}

\end{document}